\documentstyle[12pt]{article}

\begin{document}

\vskip 0.5truecm
\rightline {LNF-95/049(P)}

\vskip 2. truecm
\centerline{\bf CRITICAL BEHAVIOR OF THE SCHWINGER MODEL}
\centerline{\bf WITH WILSON FERMIONS}
\vskip 2 truecm

\centerline { V.~Azcoiti$^a$, G. Di Carlo$^b$, A. Galante$^{c,b}$,
A.F. Grillo$^d$, and V. Laliena$^a$}
\vskip 1 truecm
\centerline {\it $^a$ Departamento de F\'\i sica Te\'orica, Facultad
de Ciencias, Universidad de Zaragoza,}
\centerline {\it 50009 Zaragoza (Spain).}
\vskip 0.15 truecm
\centerline {\it $^b$ Istituto Nazionale di Fisica Nucleare,
Laboratori Nazionali di Frascati,}
\centerline {\it P.O.B. 13 - Frascati 00044 (Italy). }
\vskip 0.15 truecm
\centerline {\it $^c$ Dipartimento di Fisica dell'Universit\'a
dell'Aquila, L'Aquila 67100 (Italy)}
\vskip 0.15 truecm
\centerline {\it $^d$ Istituto Nazionale di Fisica Nucleare,
Laboratori Nazionali del Gran Sasso,}
\centerline {\it Assergi (L'Aquila) 67010 (Italy). }
\vskip 3 truecm
\centerline {ABSTRACT}
We present a detailed analysis, in the framework of the MFA approach,
 of the critical behaviour of the
lattice Schwinger model with Wilson fermions
on lattices up to $24^2$, through the study of the  Lee-Yang zeros
and the specific heat.
We find compelling evidence for a  critical line
ending at $\kappa = 0.25$ at large $\beta$.
Finite size scaling analysis on lattices $8^2,12^2,16^2, 20^2$
and $24^2$ indicates a continuous transition.
The hyperscaling relation is verified in the explored $\beta$
region.
\vfill\eject

\leftline{\bf 1. Introduction}
\par

Although apparently far from physical reality, the Schwinger model,
i.e., the theory describing the interaction between
photons and electrons in $1+1$ dimensions, has been a favourite
framework for
theoretical and numerical exercises for at least two reasons.

First, the massless Schwinger model can be analytically
solved in the continuum and this is the reason why it has always
been used as a
laboratory for the development of
numerical algorithms for dynamical fermions.

Second,
and due to the special low dimensionality dynamics, this model shares
many physical properties with four dimensional $QCD$, the gauge theory
describing the strong interaction of hadrons. In fact the charge is
confined
in the Schwinger model and long
range forces are absent in it. Of course
the Schwinger model has also some characteristic
features which are not
common to $QCD$ like the property
of superrenormalizability or
ultraconfinement (it effectively describes
free bosons) and the absence of
the Goldstone boson. However
we expect a phase diagram in the lattice
regularized Schwinger model qualitatively
similar to that of lattice $QCD$,
in particular since, as can be
derived from dimensional arguments, the
continuum limit is reached at infinite gauge coupling.
This is one of the motivations for this paper.

The Schwinger model on the lattice, using the Kogut-Susskind
regularization
for the fermion fields, has been the
subject of extensive analysis.
The phase structure
of the model in the gauge coupling
$\beta$ and fermion mass $m$ plane is
well known and the continuum value
for the chiral condensate has been
reproduced within three decimal places \cite{SCH},
the best numerical result to our
knowledge, using the $MFA$ approach \cite{MFA}.
In the Wilson regularization for
fermion fields however, the situation is
not so clear. In this scheme,
chiral symmetry is explicitly broken even for vanishing
bare fermion mass and this is the
unavoidable price to pay to overcome
the species doubling problem. Therefore
no order parameter can be used
for analyzing the phase diagram, which
besides the larger number of
degrees of freedom if compared with
the Kogut-Susskind regularization,
makes the analysis far from trivial.

Furthermore almost all the standard
algorithms to simulate numerically
dynamical fermions, like hybrid Monte
Carlo, which are based on the inclusion
of the fermion determinant in
the integration measure, need to duplicate
the number of fermion species in order
to avoid negative values for the
determinant of the Dirac operator, which
makes impossible to investigate
the phase structure of the one flavour
Schwinger model. This is the reason
why until very recent time, when an
analysis of the phase diagram of the
Schwinger model in the Wilson regularization
using non standard methods
was performed \cite{LANG}, very little was known about it.

In this article we will report the
results of a numerical simulation of the
one flavour Schwinger model in the
Wilson regularization, by means
of $MFA$ simulations. The structure of
the Lee-Yang zeroes
in the complex hopping parameter $\kappa$
plane strongly suggests
the existence of a critical line:
in our simulations the critical $\kappa$ is $0.27$ at
 $\beta=2$, ending at $\kappa=0.25$, $\beta \to \infty$,
the critical point where the continuum limit
is recovered. We also describe results
concerning the specific heat
showing a sharp peak along a line
$\kappa_c = f(\beta_c)$ in the
$\kappa, \beta$ plane, in good numerical
agreement with predictions from
the Lee-Yang zeros analysis. The results
of the finite size scaling analysis
for the Lee-Yang zeros and
chiral susceptibility allow the determination
of the critical exponents, and the fulfillment of
the hyperscaling relation gives a definite
confirmation that we are dealing with a real
second order phase transition line.

\vskip 1truecm
\leftline{\bf 2. Theoretical Grounds}
\par

The lattice action for the massive
Schwinger model with Wilson fermions
is given by

$$
S(\beta,\kappa)= S_F(\kappa) + \beta S_G
\eqno(1)$$

\noindent
where $S_F(\kappa)$ contains the kinetic
and mass terms for the fermion
field
as well as the fermion-gauge interaction term,
and we have chosen for the
pure gauge action $S_G$ the standard
noncompact regularization for the
abelian
model. The fermionic action $S_F(\kappa)$ reads

$$
S_{F}(\kappa)=\kappa \sum_{\mu,x}\,[\,\bar\psi(x+\mu)
(1+\gamma_{\mu}) U_\mu^\dagger (x)\psi(x)$$

$$
+ \bar\psi(x) (1-\gamma_\mu)
U_{\mu}(x)\psi(x+\mu)\,]\,-\,\sum_x
\bar{\psi}(x)\psi(x) \eqno(2)$$

Equation (2) defines the Dirac matrix
operator $\Delta$ which can be written
as

$$\Delta = -I + \kappa M
\eqno(3)$$

\noindent
where $I$ is the unit matrix and $M$ a
matrix of dimension equal to
the lattice volume times the number of Dirac components.

The use of the $MFA$ approach in the
numerical simulations of the system
described by action (1) has at least two
very important advantages when
compared with other standard approaches.
First, since $MFA$ is based on the
computation of the fermion effective
action defined as the logarithm of
the mean value of the fermion determinant
at fixed pure gauge energy and
this mean value is positive definite
at least in the physically
interesting region, we have no problems
to simulate the one flavour model.
Secondly, since the $\kappa$ dependence
factorizes in the non trivial part
of the fermionic operator (3),
we do not need to repeat the numerical
simulations when changing the value
of the hopping parameter $\kappa$,
which is mandatory for exploring the zeros of the
partition function in the complex $\kappa$
plane. Let us notice also that,
while $MFA$ has been extensively used
and checked in lattice gauge theories
with Kogut-Susskind fermions, this is
the first time it is applied to
a lattice model with Wilson fermions.

The technical details of $MFA$ can be
found by the interested reader in
refs. \cite{MFA}. As stated before, all
the applications of $MFA$ to
lattice gauge theories in these references
were done using staggered
fermions. The main difference between
staggered and Wilson fermions
comes from the different structure that
the Dirac operator has in each
formulation. In the Wilson case, where
the fermion matrix has the
structure given by equation (3), the
$MFA$ approach works as follows:
first we generate well decorrelated
gauge field configurations at fixed noncompact gauge
energy

$$E = {1\over{V}}\sum_{x,\mu<\nu} (A_{\mu}(x) + A_{\nu}(x+{\mu})
- A_{\mu}(x+{\nu}) - A_{\nu}(x))
\eqno(4)$$

\noindent
and then we diagonalize the $M$ matrix
for each generated configuration.
Due to the non hermiticity of the matrix
we can not use
the standard Lanczos algorithm; the
eigenvalues are found using a standard
library diagonalization routine.
{}From the eigenvalues of the $M$ matrix
we can reconstruct the determinant
of the Dirac operator $\Delta$ at any
value of the hopping parameter $\kappa$
in a trivial way. The partition function
associated to the action (1) can
be written then as

$$
Z(\beta, \kappa) =\sum_{n} C_{n}(\beta) \kappa^{n}=
\int dE n(E) e^{-\beta V E} \sum_{n} C_{n}(E) \kappa^{n}
\eqno(5)$$

\noindent
where $n(E)$ is the density of states
at fixed pure gauge energy and
$C_n(E)$ in (5) stands for the mean
value of the nth coefficient of the
polynomial describing the fermionic
determinant, the average being computed
over gauge field configurations at
fixed pure gauge energy, i.e.,

$$
C_n(E) = \int [dA_{\mu}] C_{n}(A_{\mu}(x)) \delta(\sum_{x,\mu<\nu}
(A_{\mu}(x) + A_{\nu}(x+{\mu}) - A_{\mu}(x+{\nu})
- A_{\nu}(x))-VE)
\eqno(6)$$

Since the density of states at fixed pure gauge energy

$$
n(E) = \int [dA_{\mu}] \delta(\sum_{x,\mu<\nu}
(A_{\mu}(x) + A_{\nu}(x+{\mu}) - A_{\mu}(x+{\nu})
- A_{\nu}(x))-VE)
\eqno(7)$$

\noindent
can be analytically computed in the
noncompact model, we can reconstruct
the partition function (5)
from the knowledge of the coefficients $C_n(E)$
through interpolation and one-dimensional integration.
Had we used the more
standard compact Wilson formulation for
the gauge fields, the standard
procedure would be very similar with
the only difference that in such a case,
the density of states, in general,
would have to be computed numerically.

If on the other hand we are interested
in the computation of vacuum
expectation values of physical operators
like the chiral condensate,
the standard procedure is the one
described in \cite{MFA}. We will not
repeat here the details of this
procedure but only will remember that it
is based on the computation of the
fermion effective action as a
function of the gauge energy and the
computation of the mean value of the
operator times the fermion determinant
over gauge field configurations
of fixed pure gauge energy.

\vskip 1truecm
\leftline{\bf 3. The Lee-Yang zeros in the complex $\kappa$ plane.}
\par

The first step to the determination
of the position of the Lee-Yang zeros
in the complex $\kappa$ plane is the
computation of the coefficients $C_n(E)$
of the averaged determinant of the Dirac operator.
For the calculation of $C_n(E)$ we
proceed as follows: first we chose a set
of values of energy, in the range selected
to cover the support of the
weight function in (5) for the values of
$\beta$ we are interested in.
Then for every value of $E$ in the
set we generate gauge field
configurations using a microcanonical code;
the generation of gauge fields at fixed energy
is not the costly part of the whole procedure,
so we can well decorrelate the configurations
used for measuring the fermionic
operator. Then, as stated before, we compute
exactly the eigenvalues of the
$M$ matrix from which we reconstruct by a
recursion formula the coefficients
of the fermionic determinant.

At the end we have the coefficients $C_n(E)$ evaluated at
discrete energy values: a polynomial interpolation
allows the reconstruction at arbitrary  values of the energy
$E$, in order to perform the numerical integration
in (5) and obtain the coefficients $C_n(\beta)$ that
can be regarded as the final product of this
part of the numerical procedure
for the determination of the Lee Yang zeros.
These coefficients are then
used for  the determination of the roots
of the polynomial $Z(\beta, \kappa)$.

The main features of $Z(\beta, \kappa)$ are the following:
in a volume V, the partition function is
a polynomial of order $N=2V$ in $\kappa$
and the typical range of the coefficients is of order $e^{V}$.
Thus, the main numerical problem is the efficiency of
standard root finders in the determination of the zeros.
Here we have used a method \cite{ZERI}
developed in order to analyse the
partition function zeros in the four
dimensional compact $U(1)$ model.
This algorithm is based on well known properties of analytic
functions on the complex plane which, in particular, allows the
determination
of the number of zeros for a given function inside a region of the
complex plane, provided the function has no singularities inside
this region.

In Figure 1 we plot the location of all the zeros (in the complex
$\kappa$ plane)
in a $16^2$ lattice at $\beta=7$. Figures
2 a,b contain the zeros closest
to the real axis in all the lattices used at $\beta=7,~10$

To estimate the statistical errors on the position of the
Lee-Yang zeros we followed the procedure described in \cite{ZERI}:
a standard jack-knife method is used
to produce $n$ averaged partition
functions and $n$ estimates of the location
of a given zero. The largest error on
the distance of critical zeros
(those with smallest imaginary part)
with respect to the free fermion critical point (0.25,0.0)
is of order $2\%$.
In all the cases reported in these
figures, we can see how the nearest
zero to the real axis approaches it
with increasing lattice size, thus
suggesting the existence of a phase
transition line in the $\beta-\kappa$ plane.

In Figure 3 we show the phase diagram in
the $\beta-\kappa$ plane obtained
from $8^2, 16^2$ and $24^2$  lattices.
The real part of the location of
the zero lying nearest to the
real axis gives an estimate of the
value of the critical $\kappa$.
Note that the critical line moves upward
with increasing volume.

A detailed analysis of the scaling behavior for
the Lee-Yang zeros and $\kappa_c$ with the
lattice size will be presented in section 5.

At smaller values of $\beta$ our procedure
has some problems: the signature is
the appearence of unphysical
real zeros at finite volume, so that
the estimation of the critical point becomes unsafe.
This behaviour is related to the large fluctuations in the
averaging procedure for the coefficients, already pointed out
in \cite{LANG}. Nevertheless, if we take the zeros with
smaller, but not zero, imaginary part as the critical
zeros we can continue the critical line down to $\beta=0$,
ending at $k_c=0.36$ for a $8^2$ lattice and $k_c=0.35$ for
a $16^2$ lattice. Note that in \cite{LANG}
an exact result for the
partition function at $\beta=0$ for a $8^2$ lattice
is reported, giving a critical point at $k_c=0.377$.

\vskip 1truecm
\leftline{\bf 4. The specific heat.}
\par
Chiral symmetry is always explicitly broken
and there is no order parameter
in this realization of the model,
so $\kappa$ cannot be identified with an
external field. We take it instead
as a temperature, defining the
the associated specific heat as
usual:

$$
C_\kappa=\frac{\partial^2 {\cal F}}{\partial\kappa^2} $$

\noindent where ${\cal F}$ is the free energy density.

The specific heat is  related to the chiral susceptibility $\chi$,
defined  as

$$
\chi = -2 \kappa^{2} {{d\over d\kappa} <\bar{\psi}(x)\psi(x)>}
\eqno(8)$$
differing only through a regular function
of $\kappa$ which does not influences
the critical behavior. The susceptibility,
which is what we have measured,
 diverges at the transition in the
thermodynamical limit, the divergence
becoming a sharp peak  on finite lattices.

The study of the position of the maximum
at different lattice sizes gives a
way to search for the phase transition
line which is numerically independent
from that presented in the previous
section.  We remind that, as the model
is not chirally invariant, the analogy
with a magnetic system fails.

Figures 4a,b contain the results for the
susceptibility (8) at two typical values
of the gauge coupling $\beta$ against $\kappa$ in
$8^2, 12^2, 16^2, 20^2$ and $24^2$
lattices.
As expected for a real phase transition,
the susceptibility shows a well defined maximum
at some critical value
$\kappa_c$, the height of these peaks
increasing with the lattice size and
eventually diverging in the infinite volume limit.
If we define
the critical $\kappa_c$ at each $\beta$ as
the value at which the
susceptibility $\chi$ takes its maximum
value, we get for the  lattices analyzed
the phase diagram reported in Figure  5, in
very good agreement, in the largest lattices, with
the one obtained from the analysis of the Lee-Yang zeros.

\vskip 1truecm
\leftline{\bf 5. Finite size scaling analysis.}
\par

We will now present a detailed analysis
of the scaling behavior for
the Lee-Yang zeros and specific heat,
mainly based on simulations
on $16^2, 20^2$ and $24^2$ lattices.

In the analysis of the phase diagram from the location
of the Lee-Yang zeros of Figure 3,
the real part of the zero lying next to the real axis
defines the critical $\kappa$ at each lattice size. In order to
have a real phase transition, the
imaginary part of the critical zero
should vanish in the infinite volume limit.
This has been explicitly
checked by assuming that the imaginary
part of the critical zero $z_c(L)$
as a function of lattice size is described by the function

$$
{\cal I}m~z_c(L) = a_{0} + a_{1}L^{-{1\over\nu}}.
\eqno(9)$$

In all the $\beta$ region explored,
the value of $a_0$ is compatible with zero.
The scaling analysis at this point is made using the relations
\cite{ITZY}

$$
{\kappa_c(L)-\kappa_c(\infty)} \sim L^{-{1\over\nu}}
$$

$$
{\cal I}m~z_c(L) \sim L^{-{1\over\nu}}.
\eqno(10)$$

\noindent
for the real and imaginary parts of the critical zero.

In Figure 6 we show the typical scaling
behavior as a function of the volume for the imaginary part of the
critical zero, at  $\beta= 10$.
Similar results have been obtained at different $\beta$.
{}From the slopes of the  lines
fitting the data from $16^2,~20^2,~24^2$
we determine the scaling
exponent $\nu$ as a function of $\beta$.
This is shown in Figure 7. The value of the critical index
 clearly indicates that
the transition is continuous at any value of $\beta$.
We note, moreover, that our results are compatible with
$\nu={2 \over 3}$, at any $\beta$.

Concerning the susceptibility $\chi$,
the critical $\kappa$ reported in
Figure 5 was defined, as stated before,
as the value of $\kappa$ at which
the susceptibility takes its maximum
value. Both the height $H(L)$
of the peak as well as its position
$\kappa_c(L)$ depend on the lattice size $L$.
Standard finite size scaling theory tell
us that these quantities scale
with the lattice size as \cite{FISH}

$$
H(L)\sim L^{{\alpha\over\nu}}
\eqno(11)$$
where $\alpha$ in (11) is the specific heat exponent.

In Figure 7 we present also
the scaling index of $H$ as a function of $\beta$.
Its value is again consistent with a
continuous phase transition. The
value obtained implies  $\alpha \sim \nu$.

Since in the critical region the singular part of the free
energy behaves as:

$$
{\cal F}_{sing} \sim (\kappa-\kappa_c)^{2-\alpha} \eqno(12)$$

\noindent
and given
that the only relevant length in this region
should be the correlation length, we get

$$ {\cal F}_{sing} \sim \xi^{-d} \sim (\kappa-\kappa_c)^{d\nu}
\eqno(13)$$

\noindent
from which the standard hyperscaling relation
between the correlation length
and specific heat exponents follows

$$d\nu=2-\alpha \eqno(14)$$

Our data for the critical exponents satisfy relation (14)
within the errors, in the whole $5<\beta<12.5$ region;
the typical deviation from (14) is of the order of
$0.07$.

We can conclude from the finite size scaling analysis of the
Lee-Yang zeros and chiral susceptibility
that the lattice Schwinger model
with Wilson fermions has a real continuous phase transition
ending at $\beta_c=\infty, \kappa_c=0.25$. We are able to
derive the critical exponents of the model
and we find perfect agreement with the hyperscaling relation.

In Figure 8 we show the scaling behavior of the real part of the
critical zero at $\beta=10$. The infinite
volume value of $\kappa_c$ can
be inferred using the Finite Size Scaling in (10). In
principle this analysis, carried
out for several values of $\beta$, could give an
estimate of the infinite volume critical $\kappa$ as a function of
the gauge coupling constant. However it
seems very difficult to get reliable
values for $\kappa_c(\infty)$ from the relation
(10) for the real part of
the critical zeros \cite{LANG2} since this is
not a universal relation.

\vskip 1truecm
\leftline{\bf 6. Discussion.}
\par

In the previous sections we have
shown that the Schwinger model regularized
on a lattice and with Wilson fermions has
a continuous phase transition
in the $(\beta,\kappa)$ plane, where
the correlation length diverges.
The finite size scaling analysis shows
a correlation length exponent $\nu$
taking a value around $2/3$ along
this transition line, a result which
appears very reliable since, as
previously shown, the hyperscaling
relation (14) is always verified. This
value of $\nu$ is in contrast
with the value $\nu=1$ obtained at
the end point $\beta=\infty$ of the
transition line as well as with the
value, again $\nu=1$, at $\beta=0$
suggested by the analysis reported in
\cite{LANG2},\cite{KARSCH}. We would like to
notice that violations to Universality
in fermionic systems, probably
due to the long range forces induced
by the fermion fields, have been
previously observed in the Gauged Nambu-Jona
Lasinio model \cite{SACHA}, \cite{NOS} as well as in a
fermion-gauge-scalar model \cite{JERSAK}.

Let us finally comment on the physical
meaning of the continuous
phase transition line. In $QCD_4$  this
transition line, whose existence and location
are not as clear as here, is assumed
to be the line along which the pion is
massless. In two-dimensional
models, as well known, Goldstone bosons
are absent. Moreover, the correlation
length which diverges in our case is
that associated to the
scalar particle, not the pseudoscalar one.
Nevertheless, it is our prejudice that this
is the line along which
the fermion remains massless even if, due
to the pathologies of
two-dimensional models, massless scalars or
pseudo-scalars are absent.

\vskip 1truecm
\leftline{\bf 7. Acknowledgements.}
\par

It is a pleasure to thank H. Gausterer and
C.B. Lang for very interesting
discussions and interchange of information.

This work has been partly supported through a CICYT (Spain) -
INFN (Italy)
collaboration.

\newpage
\vskip 1 truecm
\leftline{\bf Figure captions}
\vskip 1 truecm

\noindent
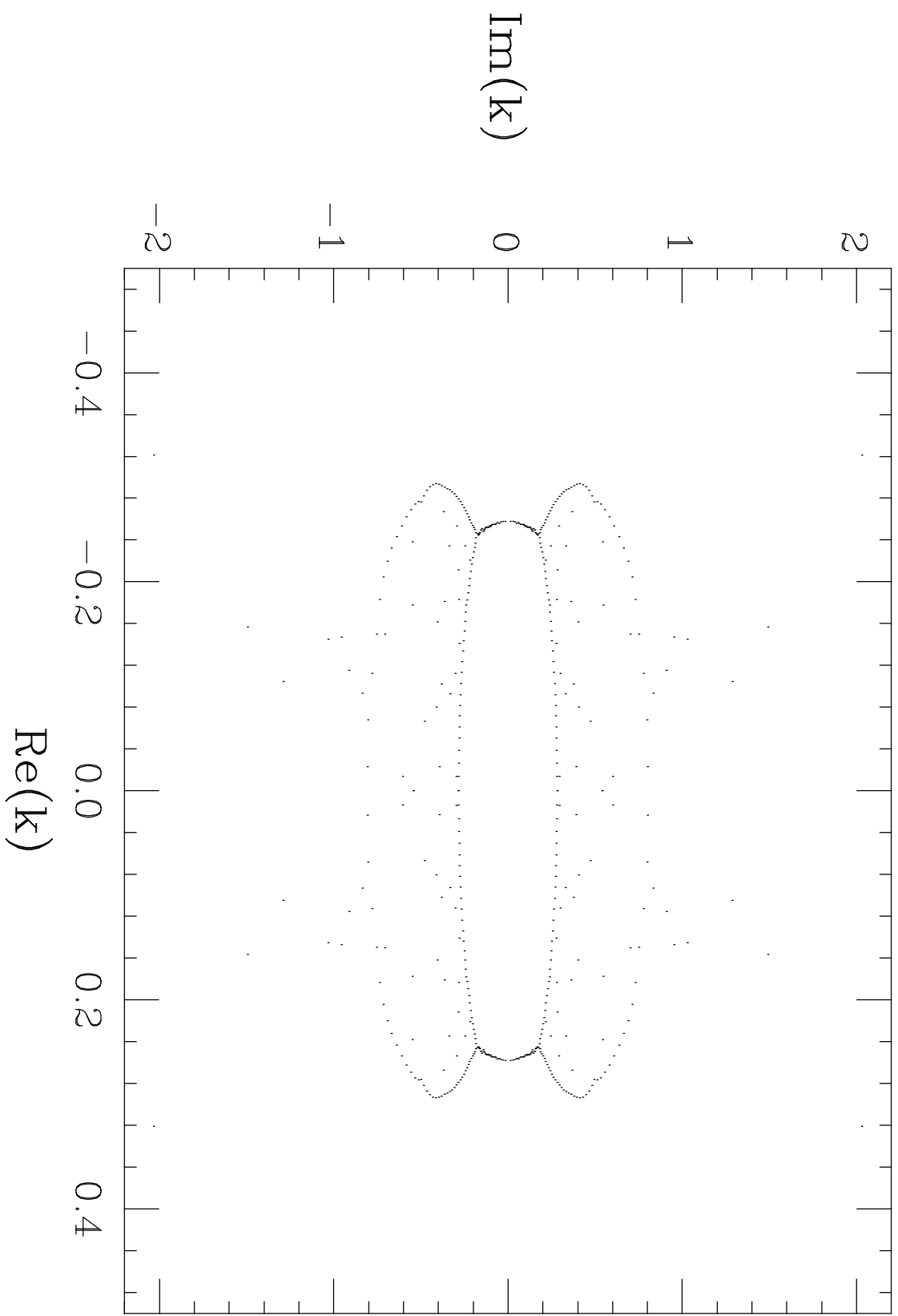
{\bf Figure 1.} Position of all the zeros at $\beta=7.0$,
$16^2$ lattice.

\noindent
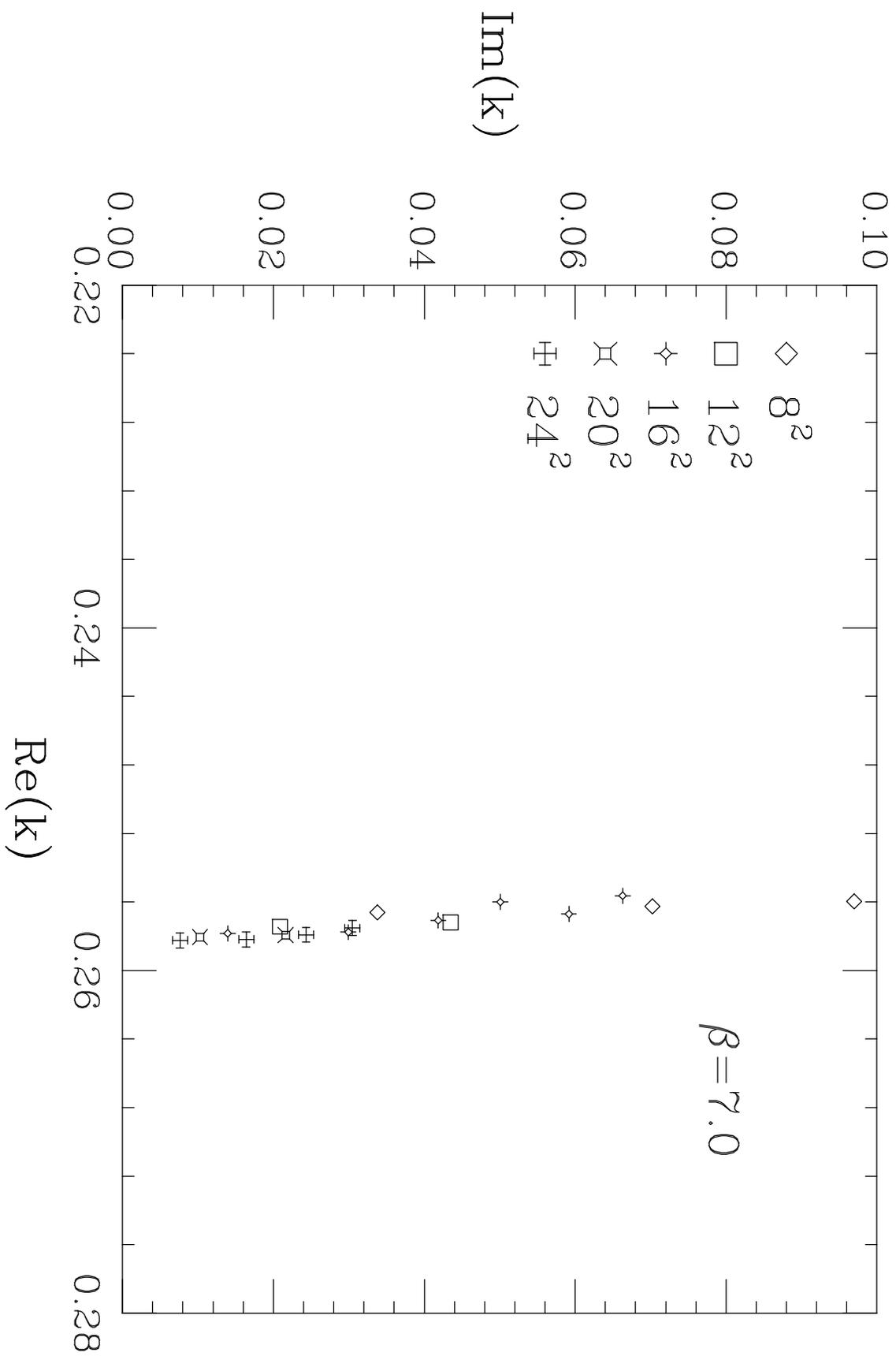
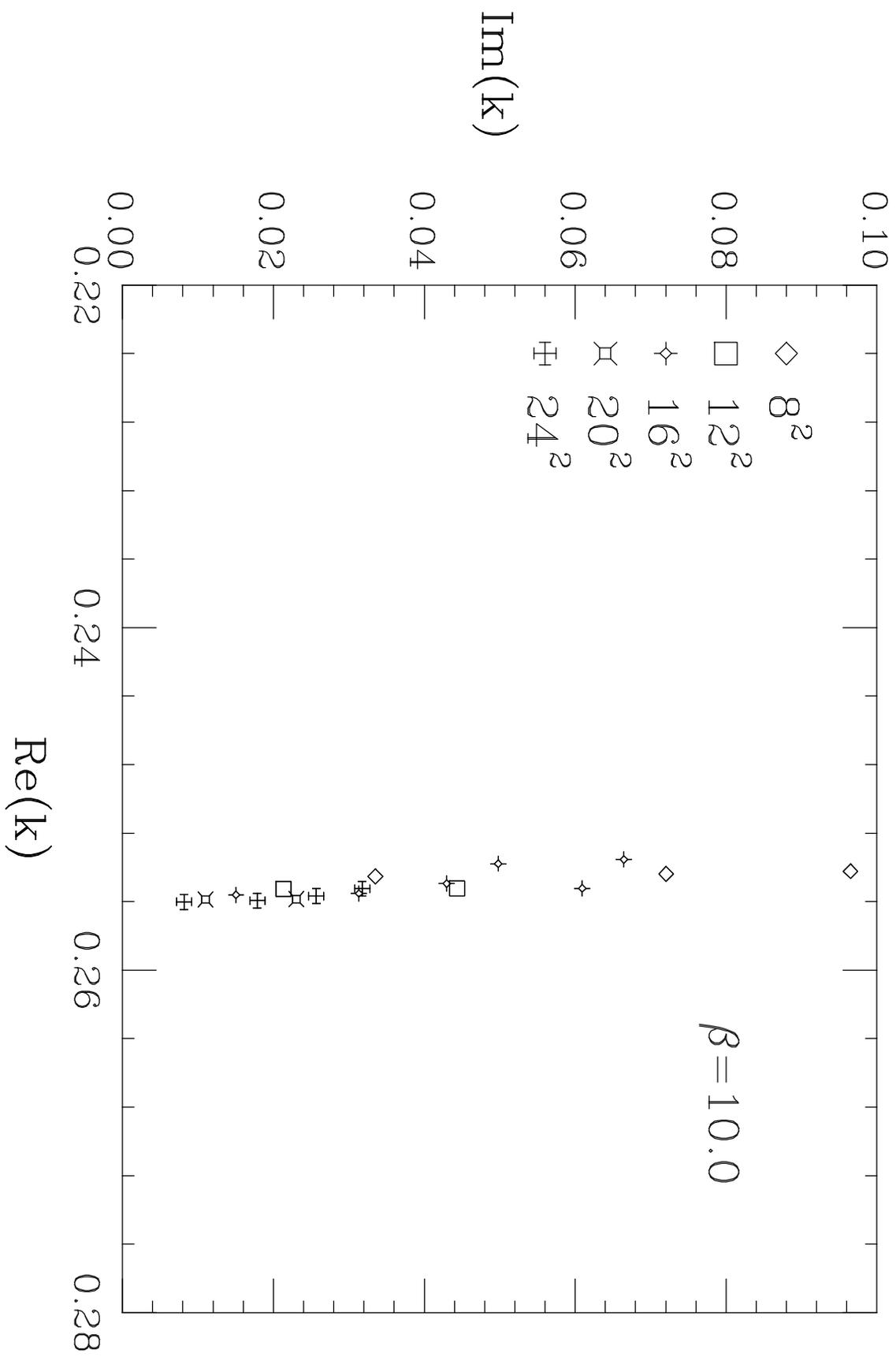
{\bf Figure 2.} Zeros closest to real axis, $\beta=7,~8^2,~12^2,
{}~16^2,~20^2,~24^2,\beta=10$ (b).

\noindent
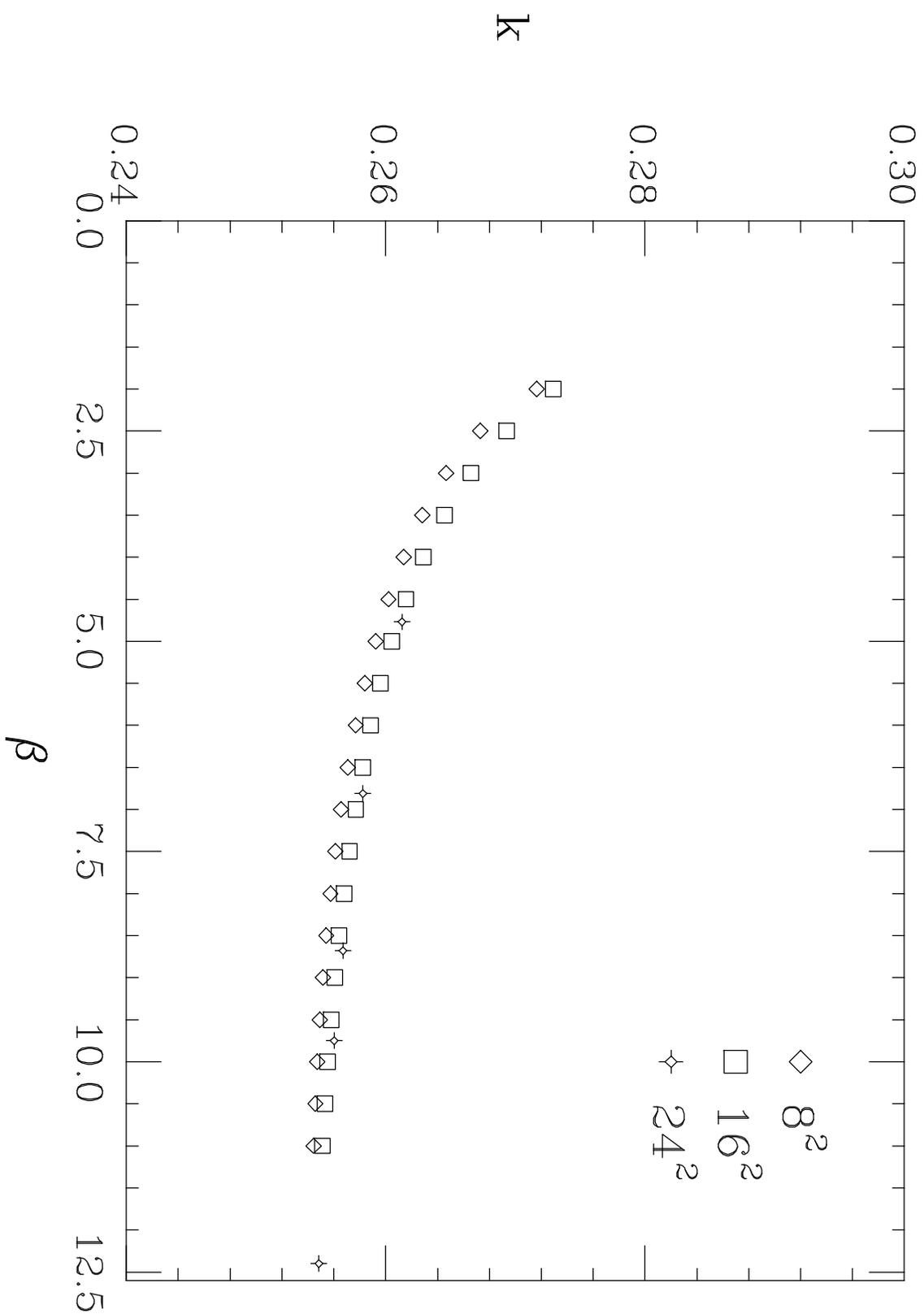
{\bf Figure 3.} $\beta-\kappa$ phase diagram from Lee-Yang zeros.

\noindent
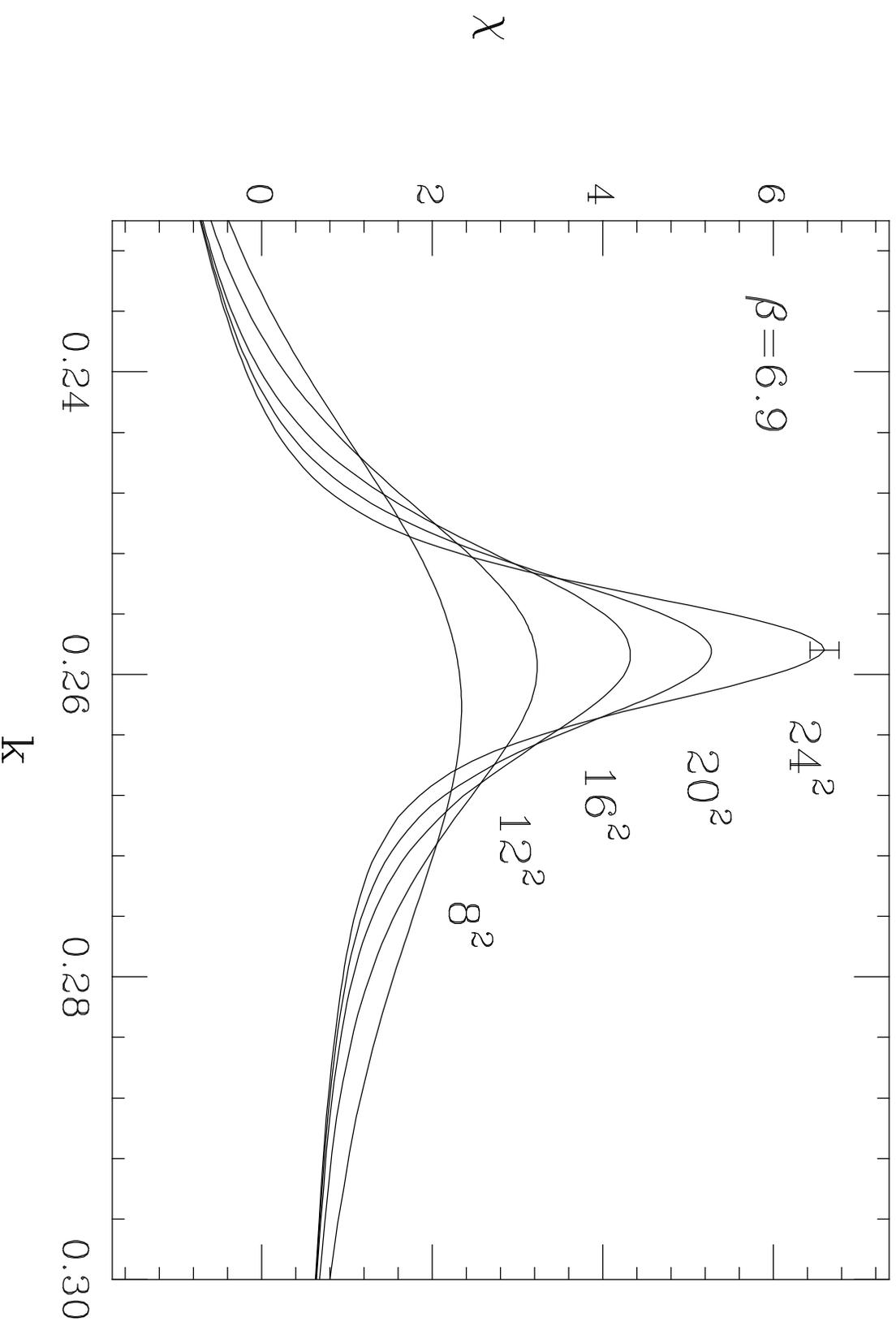
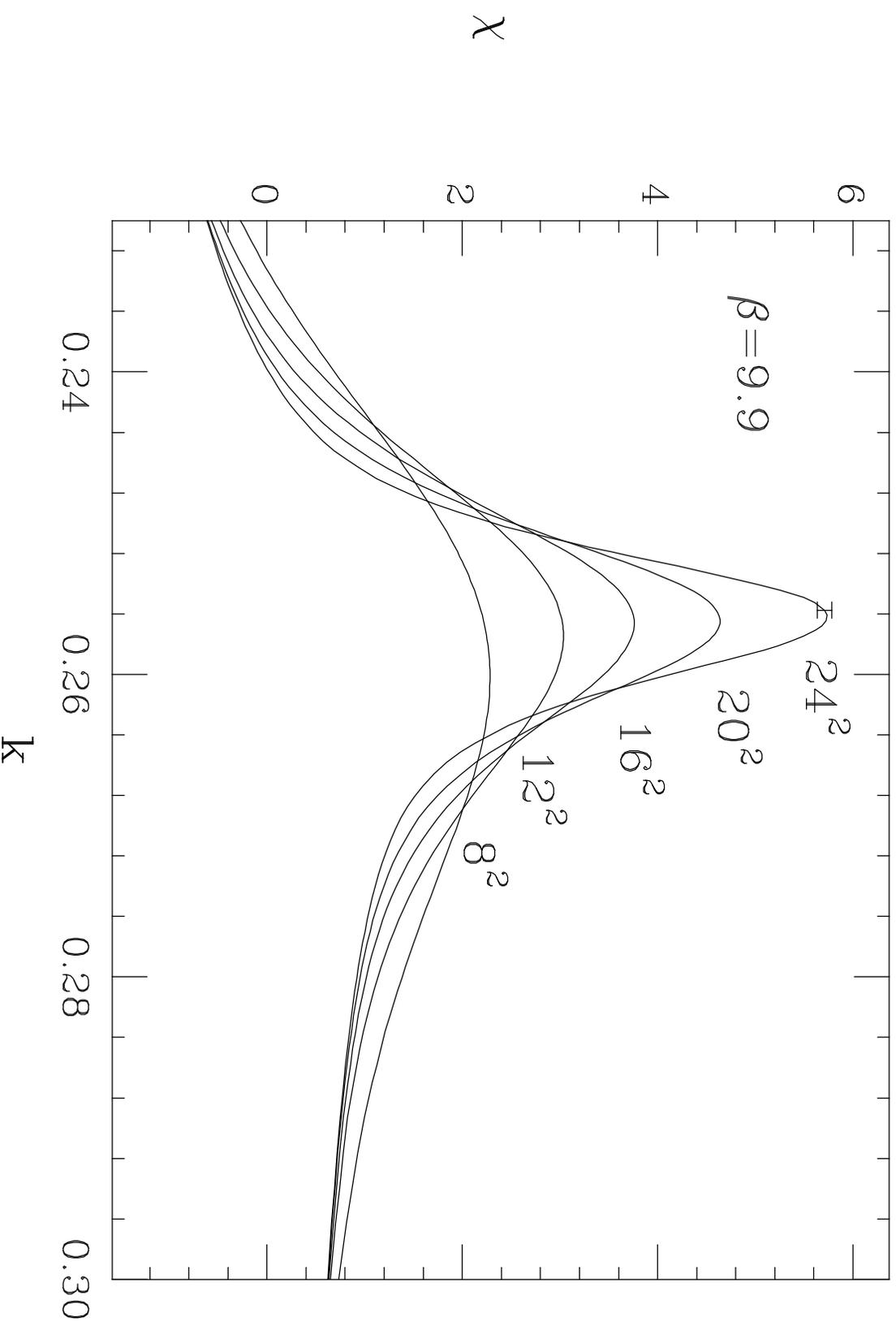
{\bf Figure 4.}  Chiral susceptibility versus $\kappa$
at $\beta=6.9$ (a) and $\beta=9.9$ (b);
the error shown is the tipical one.

\noindent
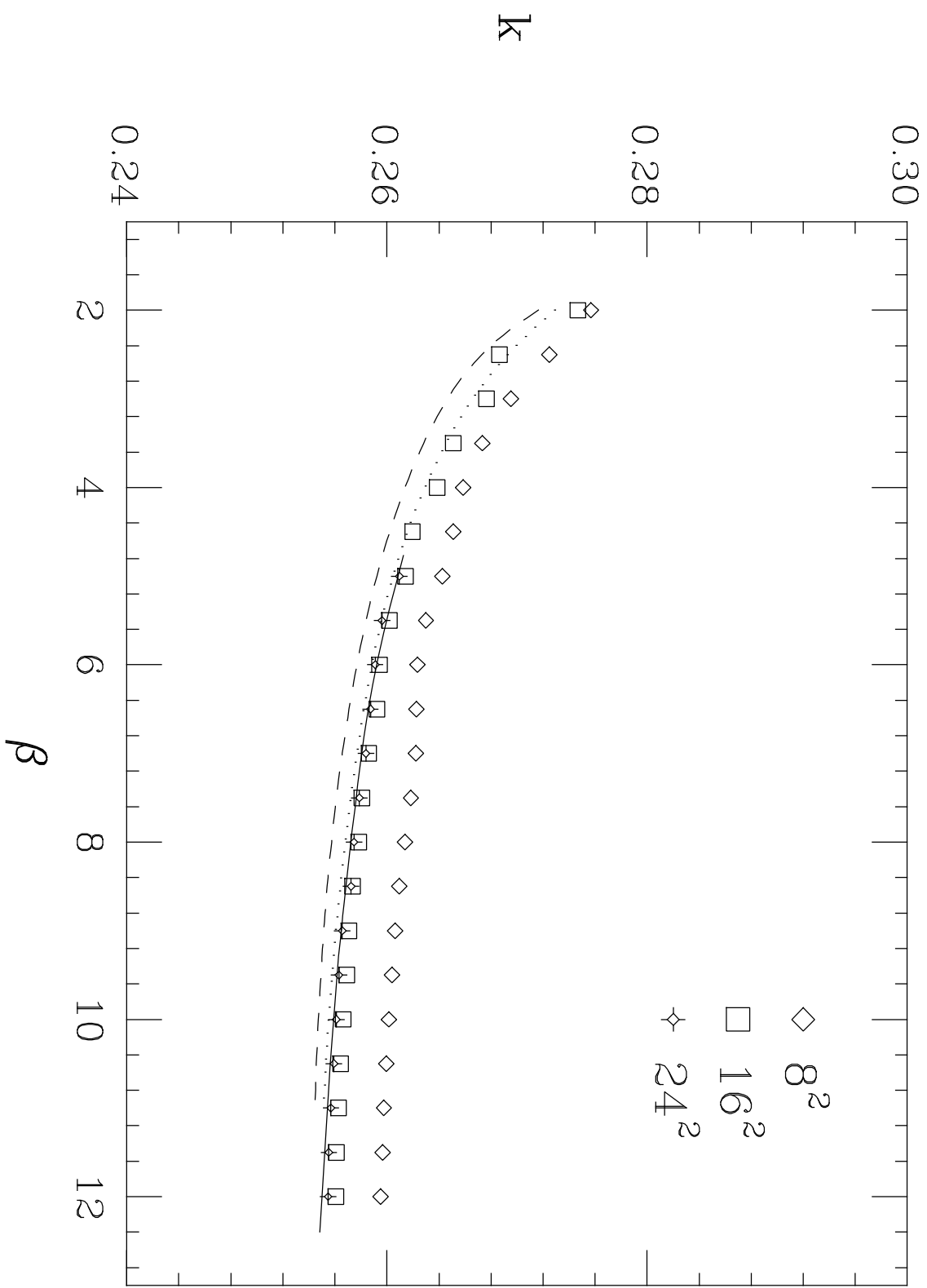
{\bf Figure 5.} $\beta-\kappa$ phase diagram from chiral
susceptibility (upper curves),
$8^2,~16^2,~24^2$ lattices, the lines are the data of Figure 3.

\noindent
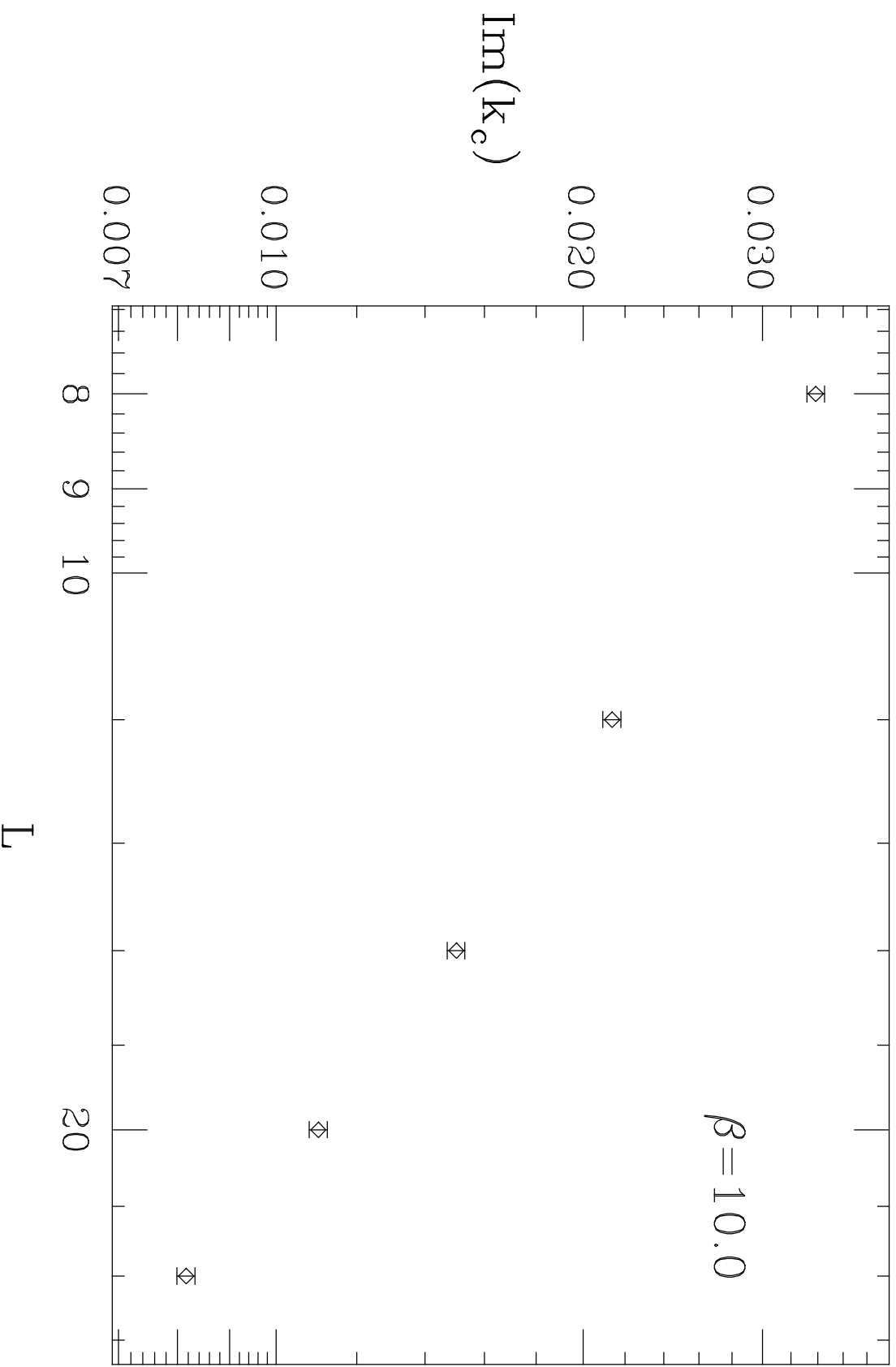
{\bf Figure 6.} Imaginary part of the critical zero versus
lattice size $L$, $\beta=10.0$.

\noindent
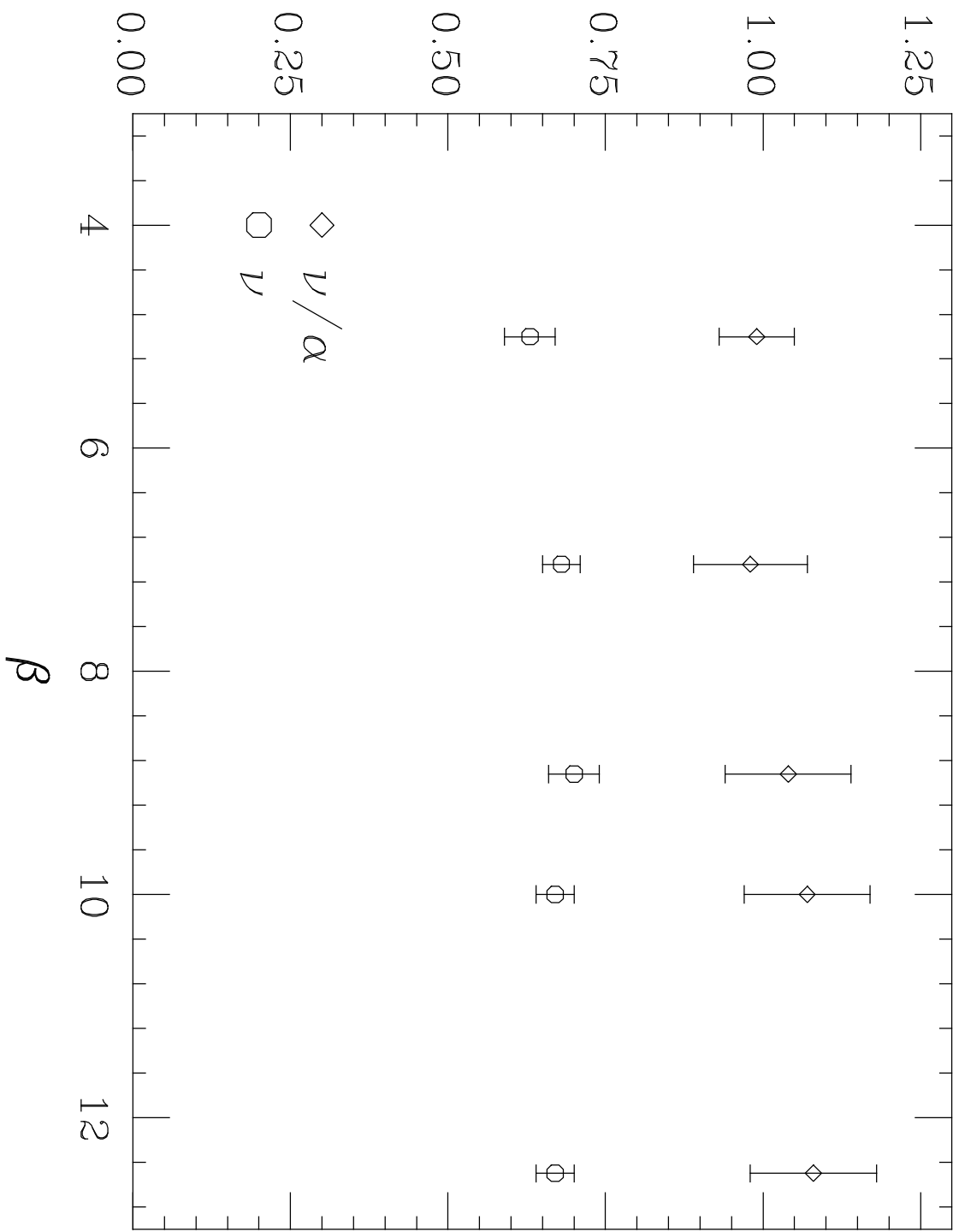
{\bf Figure 7.} $\nu$ , ${\nu \over \alpha}$
versus $\beta$, $16^2,~20^2,~24^2$ lattices.

\noindent
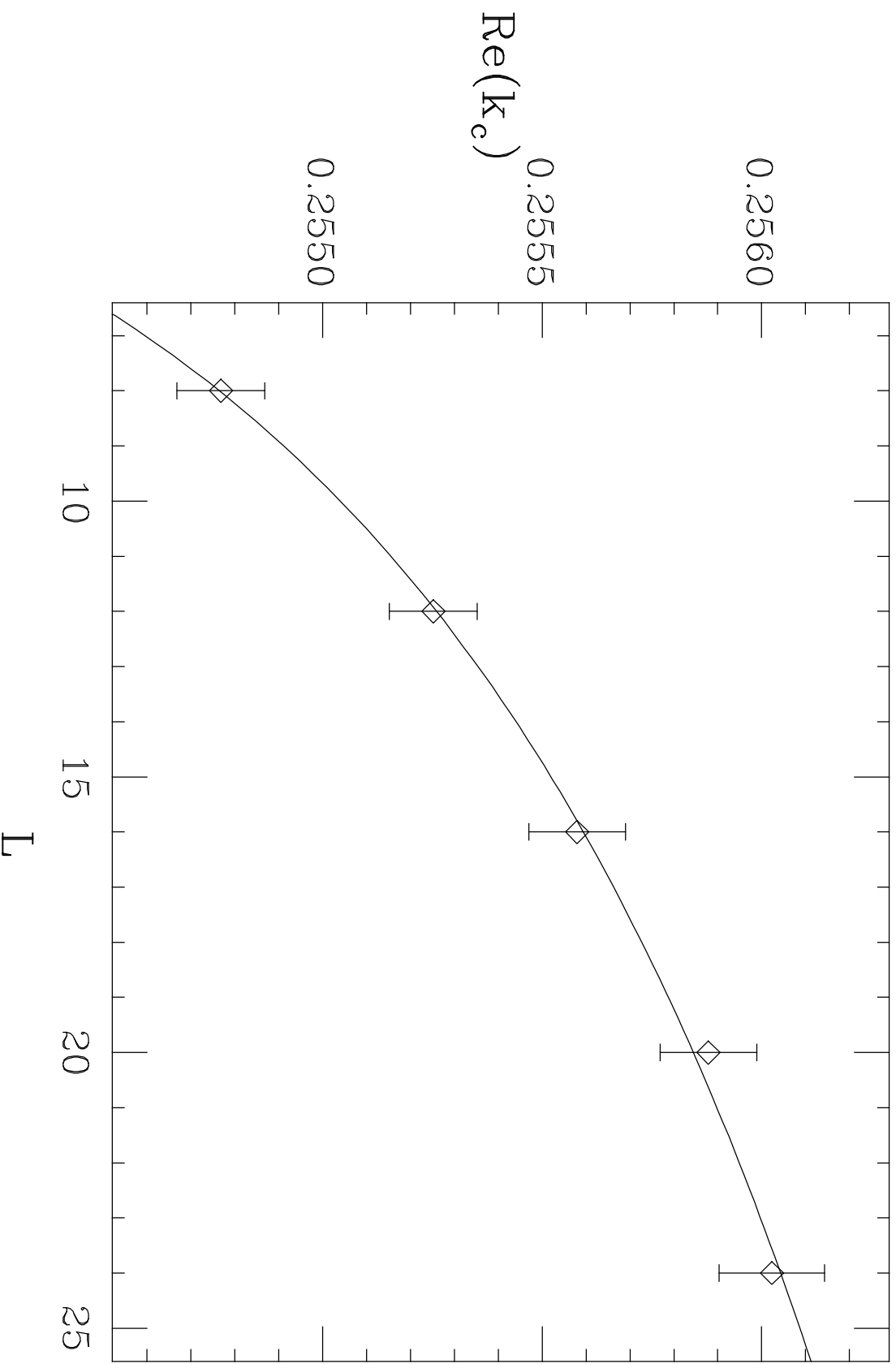
{\bf Figure 8.} Real part of the critical zero versus $L,
\beta=10.0$; the line is the best fit using (10).
\newpage

\end{document}